\documentclass[A4,aps,twocolumn]{revtex4}
 \usepackage{graphicx}
 \usepackage{color}
 \usepackage[sort&compress]{natbib}
 \usepackage{mathrsfs} 
 \usepackage{lscape,epsfig}
 \usepackage{amsmath}
 \usepackage{bm}
\usepackage{txfonts}
\usepackage{multirow}

  \def\beq{\begin{equation}}
  \def\eeq{\end{equation}}
  \def\beqa{\begin{eqnarray}}
  \def\eeqa{\end{eqnarray}}
  \def\ban{\begin{eqnarray*}}
  \def\ean{\end{eqnarray*}}
  \def\bi{\begin{itemize}}
  \def\ei{\end{itemize}}

\begin{document}

\title{Formation of hybrid stars from metastable hadronic stars} 
  \author{Domenico Logoteta, Constan\c{c}a Provid\^encia and Isaac Vida\~na}
  \affiliation{Centro de F\'{\i}sica Computacional, Department of Physics,
   University of Coimbra, PT-3004-516 Coimbra, Portugal} 
% \author{Ignazio Bombaci}
%  \affiliation{Dipartimento di Fisica ``Enrico Fermi'', 
% Universit\'a di Pisa,
% and INFN Sezione di Pisa, 
%  Largo Bruno Pontecorvo 3, I-56127 Pisa, Italy }

\begin{abstract}
We study the consequences of quark matter nucleation in cold hadronic matter  
employing three relativistic-mean-field (RMF) models to describe the hadronic phase and the 
Nambu-Jona-Lasinio (NJL) model for the quark one. We explore the effect of a vector 
interaction in the NJL Lagrangian and of a phenomenological bag constant on neutron stars metastability.  
We delineate the region of parameters of the quark phase that allow for the formation of stable hybrid stars 
with mass compatible with the almost $2 \ M_\odot$ pulsars PSR J1614-2230 ($1.97 \pm 0.04 M_\odot$) 
and PSR J0348+0432 ($2.01 \pm 0.04 M_\odot$). {It is shown, however,  that not all
  hybrid star configurations with $\sim 2 \,M_\odot$  are populated after nucleation}.
\end{abstract}
 
  \maketitle
  
  \vspace{0.50cm}
  PACS number(s): {97.60.s, 97.60.Jd, 26.60.Dd, 26.60.Kp} 
  \vspace{0.50cm}

%%%%%%%%%%%%%%%%%%%%%%%%%%%%%%%%%%%%%%%%%%%%%%%%%%%%%%%%%%%%%%%%%%%%%%%%%%%%%%%%%%%%%%%%%%%%%%

\section{Introduction}
 
%%%%%%%%%%%%%%%%%%%%%%%%%%%%%%%%%%%%%%
During the last decades the study of neutron stars has offered the possibility to investigate various  
 topics of modern physics. Due to their very large central density (several times larger 
than normal saturation density) neutron stars represent a natural observatory to study the behavior of the 
matter under extreme conditions. In this line, the issue whether neutron stars may host a deconfined quark phase 
in their cores is still an open question.

Quark matter nucleation in neutron stars has been studied by many authors 
both in cold \cite{iida98,b0a,b0,b1,b2,b3,b3a,b4,b5,b6,b7,b8,b8a,b8b} and finite temperature
\cite{h1,h2,h3,h4,me,h5,me1,h6,h7} hadronic matter, or even  in the presence of strong magnetic fields   \cite{hm1,hm2}.
These studies suggested that the nucleation process may play an important role in the emission of gamma ray bursts 
and supernovae explosions.  In most of these works, the hadronic phase was described using phenomenological 
relativistic mean field (RMF) models based on effective Lagrangian densities \cite{serot86}. 
Among the different RMF models, one of the most
popular parametrizations is that of Glendenning and Moszkowski \cite{gm91} of the 
non-linear Walecka model which has been widely used to study the effect of the 
hadronic equation of state (EOS) on the nucleation process. In particular, in Ref.\ \cite{b6} 
 the effect of different hyperon couplings on the critical mass and stellar
conversion energy was analyzed. It was found that increasing the value of the hyperon coupling 
constants, the stellar metastability threshold mass and the value of the critical 
mass increase, thus making the formation of quark stars less likely. In all these works the MIT bag model \cite{mit} 
was used to describe the deconfined phase.  In Ref.\ \cite{b9}, two models that contain explicitly the chiral symmetry were 
applied to describe the quark phase, namely the Nambu-Jona-Lasinio (NJL) model \cite{nambu}
(see also \cite{bba2,njl1}) and the Chromo Dielectric model  (CDM) \cite{cdm,cdm1}. It was shown there
that it is very difficult to populate the quark star branch using that version of the NJL model and, therefore, all compact 
stars would give pure hadronic stars in that case. On the contrary, with the CDM, both hadronic and quark star 
configurations can be formed.
Recently, in Ref.\ \cite{me3}, was discussed the possibility of quark matter nucleation using the microscopic 
Brueckner-Hartree-Fock approach to model the hadronic phase, and the three quark matter models cited above  
to describe the deconfined phase. The maximum neutron star mass predicted within this study was 
of $1.62 M_\odot$, quite far from the almost $2 \ M_\odot$ pulsars PSR J1614-2230 ($1.97 \pm 0.04 M_\odot$) \cite{Demorest10}
and PSR J0348+0432 ($2.01 \pm 0.04 M_\odot$) \cite{j0348}, recently measured.

In the present work we investigate the nucleation of quark matter in cold hadronic matter using an hadronic EOS based on three different RMF 
approaches. We consider the TM1 \cite{tm1}, the TM1-2 \cite{tm1-2} and the NL3 \cite{NL3} models. 
The TM1 and TM1-2  models satisfy the
heavy-ion flow constraints for symmetric matter around densities $2$-$3$ $\rho_0$
\cite{daniel02} (being $\rho_0=0.16$ fm$^{-3}$ the empirical
saturation point of symmetric nuclear matter). NL3, on the contrary,
does not satisfy these constraints. However it has been used  in
\cite{sedrakian} as the hadronic EOS in a scenario that allows for
hybrid stars with masses above $2$ $M_\odot$. A hard hadronic EOS seems
to be a necessary condition for the existence of massive hybrid stars. 
Although it is well known that hyperons are expected to appear in the neutron star interior
at densities $\sim 2-3 \rho_0$ and play a decisive role for several properties of
such objects, we will ignore them in this work since, as mention in the abstract, we are mostly interested in the study
of the role of the vector interaction and the phenomenological bag constant in
the NJL model, and on the determination whether the quark star branch may
  be populated.

For the quark phase we employ 
the version of the NJL model presented in Ref.\ \cite{sedrakian} but neglecting the superconducting terms. In this way 
we get an upper bound in our results, since it is generally accepted that superconductivity softens the EOS.  
In the version
of the NJL model of Ref.\ \cite{pagliara08,sedrakian}, a 
phenomenological bag
constant $B^*$ was introduced in order to define the location of the deconfinement phase transition.
A task of the present work is to delineate the region of parameters of our models that allow for the formation of stable high mass 
neutron stars after the nucleation process. 
For the formation of an hybrid star it is important that the nucleation time of the
metastable hadronic star,  from which it originates, be smaller than the age of the universe.       

%%%%%%%%%%%%%%%%%%%%%%%%%%%%%%%%%%%%%%%%%%%%%%%%%%%%%%%%%%%%%%%%%%%%%%%%%%%%%%%%%%%%%%%%%%%%%%%%%%%%%%%%%%%%
 \section{The hadronic equation of state}
\label{sec:EOS}

\begin{table}
\begin{ruledtabular}
\begin{tabular}{cccccccccccc}
Model & $\left(\frac{g_{\sigma}}{m_{\sigma}}\right)^2$ & $\left(\frac{g_{\omega}}{m_{\omega}}\right)^2$ & $\left(\frac{g_{\rho}}{m_{\rho}}\right)^2$  & $b$ & $c$ & $\xi$\\
 & $(\hbox{fm})^2$ & $(\hbox{fm})^2$ & $(\hbox{fm})^2$ &  & & \\
\hline
%TM1    &  15.0125 & 10.1187 & 5.6434  & 3.0655 & 2.7333   & 0.016  \\
%TM1-2  &  14.9065 & 9.9356 & 5.6434   & 3.5351 & -47.8812 & 0.011  \\
NL3    &  15.737  & 10.523  & 1.338    & 0.002055 & -0.002651 & 0.0       \\
TM1    &  15.0125 & 10.1187 & 5.6434   & 0.001450 &  0.000044 & 0.016  \\
TM1-2  &  14.9065 & 9.9356  & 5.6434   & 0.001690 & -0.000797 & 0.011  \\
\end{tabular}
\end{ruledtabular}
\caption{{Coupling constants for the NL3, TM1 and TM1-2 models. For the
    TM1 and TM1-2 models the value of $\Lambda_\omega=0.03$ ($L=55$ MeV) has
    been considered while for the NL3 model no $\omega$-$\rho$ has been
    included being therefore $\Lambda_\omega=0$  ($L=118$ MeV) in this case.}}
\label{coeff}
\end{table}
As said before, in this work we have used three popular relativistic mean filed models to describe the hadronic phase of our system, namely the 
NL3, the TM1 and the TM1-2 models. These models are based on the following Lagrangian density: 
\begin{eqnarray}
\mathcal{L}&=&\sum_N \bar{\psi}_N[\gamma^{\mu}(i\partial_{\mu}-g_{\omega N}\omega_{\mu
}-\frac{1}{2} g_{\rho N} {\boldsymbol \tau} \cdot {\boldsymbol \rho_{\mu}} )\nonumber \\
&-&(m_{N}-g_{\sigma N}\sigma)]\psi_N+\frac{1}{2}\partial^{\mu}\sigma\partial_{\mu
}\sigma-\frac{1}{2}m_{\sigma}^{2}\sigma^{2} \nonumber \\
&+&\frac{1}{2}m_{\omega}^{2}\omega^{\mu}\omega_{\mu}-\frac{1}{4}{\boldsymbol
  \rho}^{\mu\nu} \cdot {\boldsymbol \rho}_{\mu\nu}+\frac{1}{2}m_{\rho}^{2}{\boldsymbol
  \rho}^{\mu}\cdot{\boldsymbol \rho}_{\mu}\nonumber \\
&-&\frac{1}{3}b m_{N} (g_{\sigma N} \sigma)^3
-\frac{1}{4}c (g_{\sigma N} \sigma)^4 \nonumber \\
&-& \frac{1}{4} \Omega_{\mu \nu} \Omega^{\mu \nu}
+\frac{1}{4!}\xi g_{\omega}^4 \left(\omega_{\mu}\omega^{\mu}\right)^2 \nonumber \\
&+&\Lambda_{\omega}\left(g^{2}_{\omega} \omega_{\mu}\omega^{\mu}\right)\left(g^{2}_{\rho} \boldsymbol{\rho}_{\mu}
\cdot\boldsymbol{\rho}^{\mu}\right) \nonumber \\
&+&\sum_{l=e^-, \mu}\bar{\psi}_{l}(i \gamma^{\mu} \partial_{\mu}-m_l)\psi_l \;,
\label{lagran}
\end{eqnarray}
where the sum is performed over nucleons, $\psi_N$
represents the corresponding Dirac
field, and interactions are mediated by the $\sigma$ isoscalar-scalar,
$\omega_{\mu}$ isoscalar-vector and $\rho_{\mu}$ isovector-vector
meson fields. The mesonic field tensors are given by their usual expressions: 
$\Omega_{\mu \nu}=\partial_{\mu}\omega_{\nu}-\partial_{\nu}\omega_{\mu}$,
$\boldsymbol{\rho}_{\mu \nu}=\partial_{\mu}\boldsymbol{\rho}_{\nu}-
\partial_{\nu}\boldsymbol{\rho}_{\mu}$. 
The values of nucleon-meson couplings and the other parameters of the Lagrangian are reported in Table\ \ref{coeff}. 

In this work we have included just nucleons in the hadronic phase. 
 
We note, however, that hyperons are expected to appear in neutron star matter at densities of $2-3 \ \rho_0$.
Their presence in neutron stars has been studied by many authors using either phenomenological
\cite{gm91,phen} and microscopic \cite{micr} approaches since the pioneer work of Ambartsumyan and Saakyan \cite{ambart60}.
It is well known that their appearance softens the EOS leading to a substantial reduction
of  the neutron star mass. Recently, it has been shown that
the inclusion of mesons with hidden strangeness and, particularly, a weak scalar coupling
and a strong vector coupling, may give rise to a quite hard EOS allowing for quite massive stars with hyperonic degrees of freedom
\cite{schaffner2012,tm1-2}. However, our present knowledge of the hyperon interactions (particularly, the hyperon-hyperon one)
is yet not very well constrained by experimental data. Therefore, the result of these works should
be taken with care.  The results of our calculation without hyperons should be interpreted just as an upper limit for
the maximum star mass. If it is not possible to get a two-solar mass neutron star including only nucleonic degrees of freedom in the
hadronic phase, then the presence of hyperons most probably will only worsen this situation.
Some results including hyperons will  be, however, shown for completeness (see discussion below).  

The NL3 model does not contain neither the quartic term in $\omega$ nor the nonlinear $\omega$-$\rho$ one. Their respective coefficients 
$\xi$ and $\Lambda_\omega$ are put to zero in Table\ \ref{coeff}. 
The NL3 model has the following saturation properties: saturation density $\rho_0=0.148$ fm$^{-3}$, binding 
energy $E/A= -16.30$ MeV, symmetry energy $J=37.4$ MeV, incompressibility $K= 271.76$ MeV and effective mass $M^*/M= 0.60$.
For the TM1 and the TM1-2 models all the terms in the Lagrangian (\ref{lagran}) are nonzero. 
The quartic term in $\omega$ was proposed in Ref.\ \cite{tm1} in order to get a RMF model able to fit the ground-state properties of several 
nuclei and Dirac-Brueckner-Hartree-Fock calculations at large densities. 
The nonlinear $\omega$-$\rho$ term is instead needed to get a good value for the slope of the symmetry energy $L$ at saturation density as suggested in Ref.\ \cite{horo01}. 
The original TM1 model, with $\Lambda_\omega=0$, 
 predicts a value of $L=110$ MeV that is too high according to the experimental constraints  
coming from different nuclear properties, lying close to the upper limit of isospin diffusion in heavy ion collisions\ \cite{chen05}. 
Taken $\Lambda_\omega=0.03$ a more reasonable value of $L=55$ MeV is obtained.
The TM1 and the TM1-2 have the same saturation properties: saturation density $\rho_0=0.145$ fm$^{-3}$, binding energy $E/A= -16.30$ MeV, symmetry energy $J=36.93$ MeV, incompressibility $K= 281.28$ MeV and effective mass $M^*/M= 0.63$. 

%%%%%%%%%%%%%%%%%%%%%%%%%%%%%%%%%%%%%%%%%%%%%%%%%%%%%%%%%%%%%%%%%%%%%%%%%%%%%%%%%%%%%%%%%%%%
\section{The quark matter equation of state}

For the description of the high-density quark matter we have employed the NJL Lagrangian, 
 extended to include the t' Hooft interaction term (proportional to $K$)  
and the vector interaction (proportional to $G_V$):  
\begin{eqnarray}
\mathcal{L}_{NJL}&=&\bar\psi(i\gamma^{\mu}\partial_{\mu}-\hat m)\psi \nonumber\\
&+&G_S \sum_{a=0}^8 [(\bar\psi\lambda_a\psi)^2+(\bar\psi i\gamma_5 \lambda_a\psi)^2]\nonumber\\
&-&K \left \{ {\rm det}_{f}[\bar\psi(1+\gamma_5)\psi]+{\rm det}_{f}[\bar\psi(1-\gamma_5)\psi]\right\} \nonumber \\
&-&G_V \sum_{a=0}^8 [(\bar\psi\gamma_\mu \lambda_a \psi)^2+(\bar\psi \gamma_5\gamma_\mu\lambda_a \psi)^2 ] \;,
\label{Lagrangian_njl}
\end{eqnarray}
where the quark spinor fields $\psi_{\alpha}$ carry
 a flavor ($\alpha= u, d, s$) index, the matrix of quark current
masses is given by $\hat m= {\rm diag}_f(m_u, m_d, m_s)$, $\lambda_a$
with $ a = 1,..., 8$ are the well known Gell-Mann matrices in the color space,
and $\lambda_0=(2/3) { 1_f}$. 
At zero temperature the pressure is given by: 
\begin{eqnarray}
p&=&\frac{1}{2\pi^2}\sum_{i=u,d,s} \int_{0}^{\Lambda}dk k^2
\vert\epsilon_i\vert-2 G_s\sum_{i=u,d,s}\sigma_{i}^2 \nonumber\\
&+& 4 K \sigma_u\sigma_d\sigma_s %\nonumber \\
-2 G_V \sum_{i=u,d,s}n^2_i-B_0-B^* \nonumber\\
&+&\sum_{l=e^-,\mu^-} p_l,
\end{eqnarray}
where $\epsilon_i$ are the quasiparticle spectra of quarks, $\sigma_i$ are quark condensates, $n_i$ quark number densities, 
$p_l$ is the lepton pressure, $B_0$ is the
vacuum pressure and $B^*$ is an effective bag constant. The quark
chemical potentials are modified by the vector fields as follow: 
$\mu_i^*=\mu_i-4 G_V n_i$.
The numerical values of the parameters of the Lagrangian are $m_{u,d}
= 5.5$ MeV, $m_s = 140.7$ MeV, $\Lambda = 602.3$ MeV, $G_S\Lambda^2 =
1.835$, $K\Lambda^5 =12.36$.

%%%%%%%%%%%%%%%%%%%%%%%%%%%%%%%%%%%%%%%%%%%%%%%%%%%%%%%%%%%%%%%%%%%%%%%%%%%%%%%%%%%%%%%%%%%%
 \section{Quark matter nucleation in hadronic stars}
\label{sec:pe}

In bulk matter the hadron-quark mixed phase begins at the ``static transition point'', defined
according to the Gibbs criterion for phase equilibrium:
\begin{equation}
\mu_H=\mu_Q \equiv \mu_0 \ , \,\,\,\,\, P_H(\mu_0)=P_Q(\mu_0)\equiv P_0 \ ,
\label{eq:p0}
\end{equation}
where
\begin{equation}
\mu_H=\frac{\epsilon_H+P_H}{n_H} \ , \,\,\,\,\, \mu_Q=\frac{\epsilon_Q+P_Q}{n_Q} 
%\label{eq:p0}
\end{equation}
are the Gibbs energies per baryon ({\em i.e.,} average chemical potentials) for the hadron (H) and
quark (Q) phases, respectively, and the quantities $\epsilon_H (\epsilon_Q)$, $P_H (P_Q)$, and $n_H (n_Q)$
denote respectively the total ({\em i.e.,} including leptonic contributions), energy density, total pressure,
and baryon number density of the two phases. The deconfinement transition in the high density region relevant for
neutron stars is assumed to be of first order.
The pressure $P_0$ defines the transition pressure. For pressures
above $P_0$ the hadronic phase is metastable, and the stable quark phase will appear as a result of a nucleation
process. The time scale of the deconfinement transition is determined by the strong interaction and, therefore,
quark flavor must be conserved during the deconfinement transition. We call $Q^*$ phase the deconfined quark matter,
in which the flavor content is equal to that of the $\beta$-stable hadronic phase at the same pressure and
temperature. Due to the weak interaction the flavor content of the deconfined droplet will soon change
after deconfinement, and a droplet of $\beta$-stable quark matter is formed. Once the first seed of quark matter
is formed the pure hadronic star will ``decay'' into an hybrid or a quark star \cite{oli87,hbp91,grb}. It was
shown in Refs.\ \cite{b0,b1,b2,b3,b4,b5,b6,b7} that pure hadronic stars with values of the central pressure, $P_c$,
larger than $P_0$ are metastable, and that their mean lifetime depend dramatically on $P_c$. As in Refs.\
\cite{b0,b1,b2}, in this work, we define the critical mass $M_{cr}$ of cold and deleptonized stars as the value
of the gravitational mass of the metastable hadronic star for which the nucleation time $\tau$ is $\sim 1$ yr.

The nucleation process of quark matter in hadronic stars can proceed both via quantum tunneling (at zero or 
finite temperature) or thermal activation \cite{me1}. In the present work we only consider cold stellar matter, and, 
therefore, nucleation only via quantum tunneling. Here we follow closely the formalism presented in \cite{iida98,b2}.

The process of formation of the drop is regulated by its quantum fluctuations in the potential
well created from the difference between the energy densities of the hadron and quark phases.
Keeping only the volume and the surface terms, the potential well takes the simple form
\begin{equation}
   U({\cal R}) = \frac{4}{3}\pi n_{Q^*}(\mu_{Q^*} - \mu_H){\cal R}^3 + 4\pi \sigma {\cal R}^2 \;,
 \label{eq:potential}
 \end{equation}
where ${\cal R}$ is the radius of the droplet, and $\sigma$ is the surface tension for the surface 
separating the hadronic phase from the $Q^*$ phase.
Within the Wentzel--Kramers--Brillouin (WKB) the quantum nucleation time is equal to
 \begin{equation}
   \tau_q = (\nu_0 p_0 N_c)^{-1} \; , 
 \label{eq:time}
 \end{equation} 
where $p_0$ is
 the probability of tunneling  given by
 \begin{equation}
   p_0=exp\left[-\frac{A(E_0)}{\hbar}\right] \;.
 \label{eq:prob}
 \end{equation}
$A(E)$ the action under the potential barrier, which in a relativistic framework reads
 \begin{equation}
  A(E)=\frac{2}{c}\int_{{\cal R}_-}^{{\cal R}_+}\sqrt{[2m({\cal R})c^2 +E-U({\cal R})][U({\cal R})-E]} \ d{\cal R} \;,
 \label{eq:action}
 \end{equation} 
being ${\cal R}_\pm$ the classical turning points, $m({\cal R})=4\pi n_H(1-n_{Q^*}/n_H)^2{\cal R}^2$ 
the droplet effective mass, $E_0$  and $\nu_0$ are the ground state energy and the oscillation frequency 
of the drop in the potential well $U({\cal R})$, respectively. In Eq.\ (\ref{eq:time}) $N_c \sim 10^{48}$ is the number of 
nucleation centers expected in the innermost  part ($r \leq R_{nuc} \sim100$ m) of the hadronic star, where pressure and 
temperature can be considered constant and equal to their central values. 
% 
%
%%%%%%%%%%%%%%%%%%%%
%
\begin{figure}[t]
 \vspace{0cm}
% \begin{tabular}{cc}
 \includegraphics[width=1.\linewidth,angle=0]{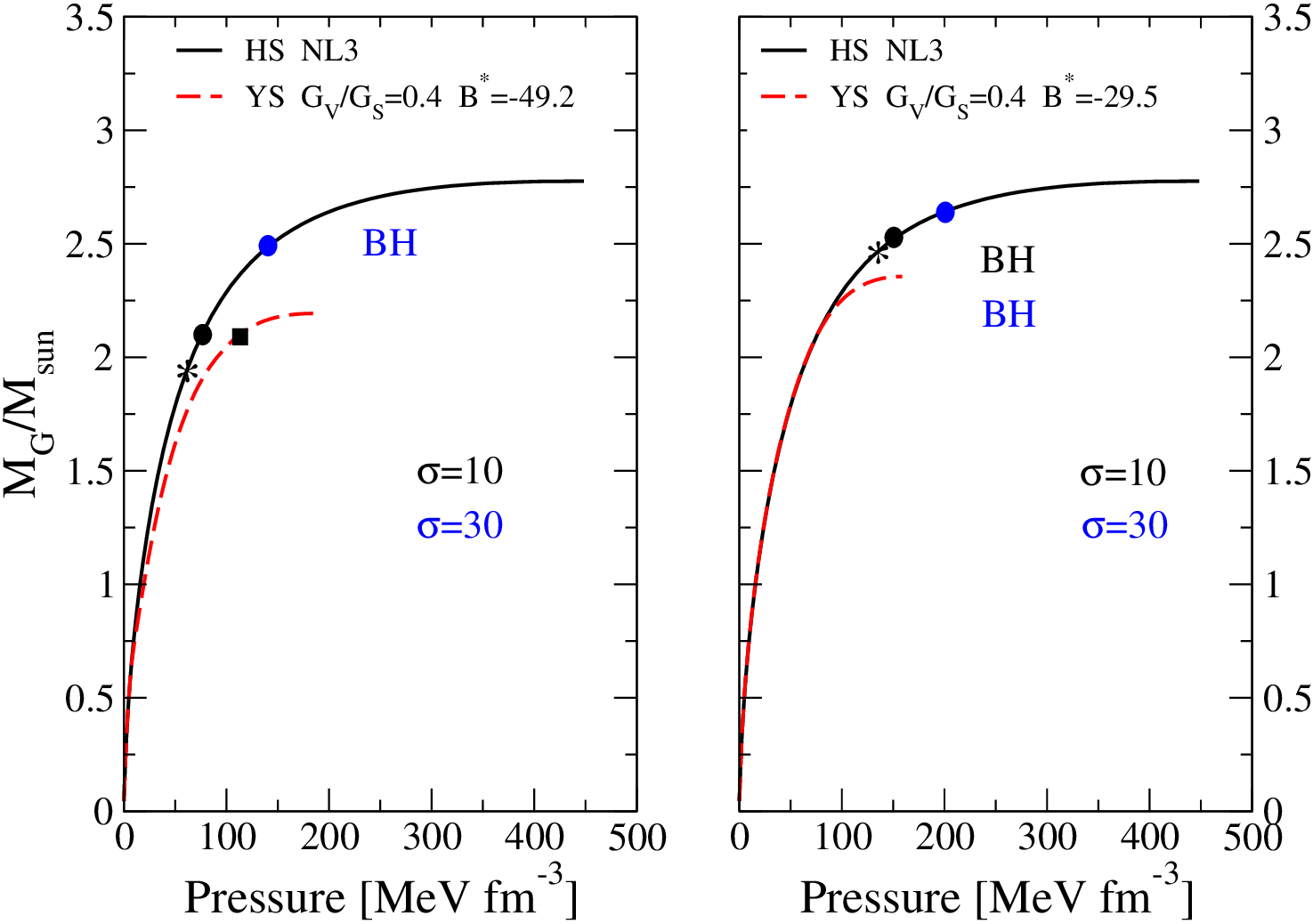}
 \includegraphics[width=1.\linewidth,angle=0]{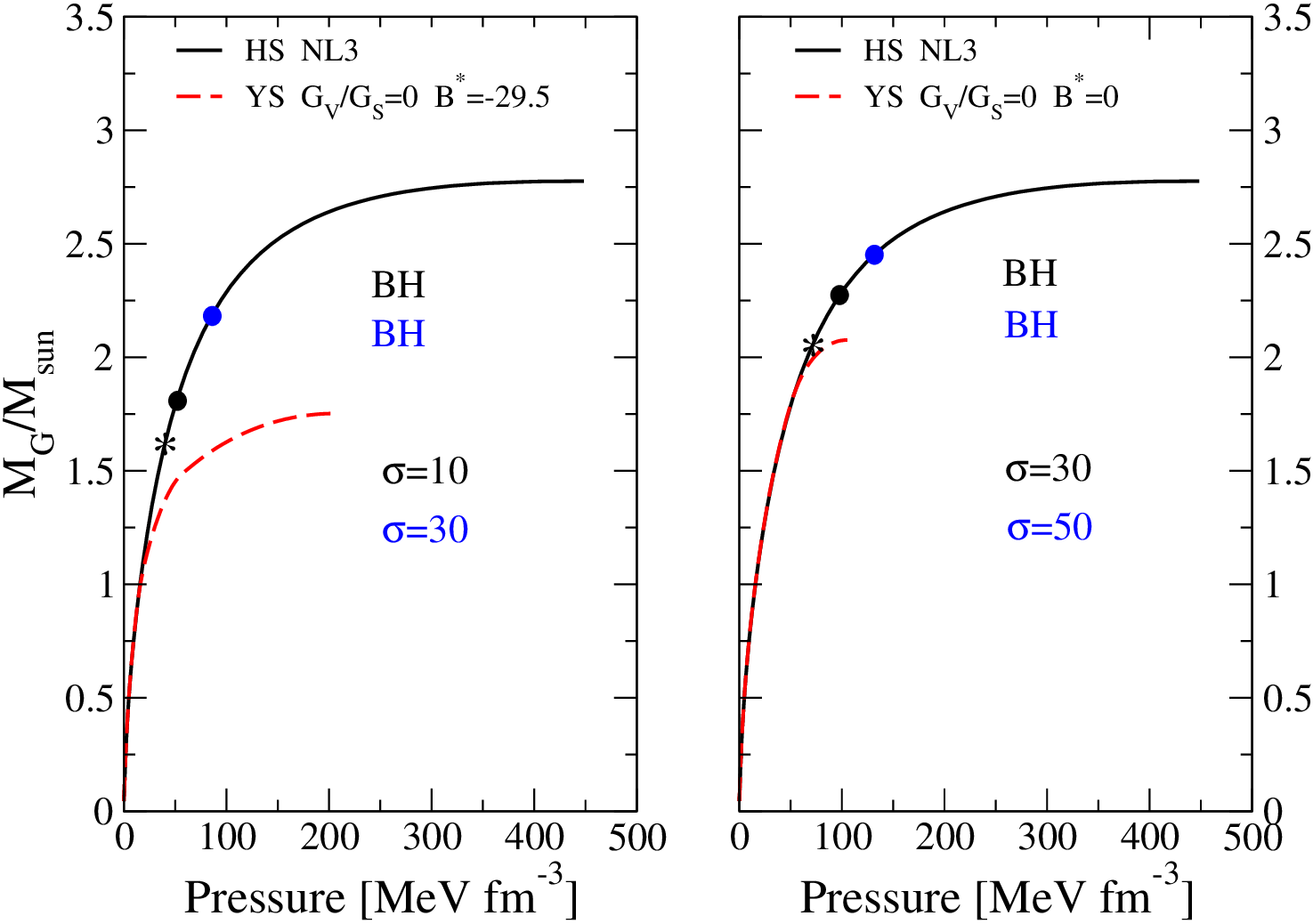}
% \end{tabular}
\caption{ (Color online) Gravitational mass versus central pressure for compact stars. Hadronic star sequences are calculated
using the NL3 parametrization for pure nucleonic matter (black curve). The hybrid star (YS) sequence is represented 
by the red dashed curve. 
The quark phase is described by the NJL model with different values of $G_V/G_S$ and $B^*$. Results are shown
for two different surface tensions. The configuration marked with an asterisk represents 
in all cases the hadronic star for which the central pressure is equal to $P_0$ and thus the quark matter nucleation time is 
 $\tau=\infty$. The critical mass configuration ($\tau=1$ yr) is denoted by a full circle. 
In panel (a) for $\sigma=10$ MeV/fm$^2$, the final quark star mass is denoted by a black square on the YS sequence. 
In the other cases reported in the figure, the quark matter 
nucleation process will lead to the formation of a black hole (BH). }   
\label{fig1}
\end{figure} 
%
%%%%%%%%%%%%%%%%%%%%

%%%%%%%%%%%%%%%%%%%%%%%%%%%%%%%%%%%%%%%%%%%%%%%%%%%%%%%%%%%%%%%%%%%%%%%%%%%%%
  
\section{Results and discussion}
\label{sec:results}

In this section we show the results of our calculations in which we have used the models previously discussed. 
Chosen a model for the hadronic part of our system, our version of the NJL model contains the free parameter $B^*$, that we have 
considered as an effective bag pressure, and the coupling of the vector interaction $G_V$. In addition, the scarce knowledge of the surface tension between the hadronic and the quark phase,  
introduces another parameter, the surface tension $\sigma$. 
Recently, the surface tension of quark matter was calculated within the two
flavor sigma model and the two- and three-flavor NJL model \cite{marcus12} and a value
in the range $7$-$30$ MeV/fm$^2$ was obtained. We will mostly use values
of $\sigma$ within these range.

  A study of finite size effects
  between the hadronic and the quark phase was also performed in several works
  \cite{endo06,maru07}. The main conclusion of these works is that for large values of the
  surface tension, namely above $40$ MeV/fm$^2$, the hadron-quark phase
  transition is closer to a Maxwell than to a Gibbs construction. However, 
there are still many uncertainties on the approach used to model the
hadron-quark phase transition and, as it was mentioned, in  \cite{marcus12} a
surface tension in the range $7$-$30$ MeV/fm$^2$ was obtained. A small surface
tension will bring the whole picture closer to the Gibbs construction.
 A wide discussion on the   
advantages and drawbacks in using a Gibbs or a Maxwell construction can be
found in the following references \cite{yasutake09,mishustin2009,Max_Gibbs}. 
We will perform the present discussion within the Gibbs construction. This
will mean that we will be able to obtain hybrid star configurations with both a
pure quark phase or  a mixed hadronic-quark phase in the star center. Within the Maxwell construction hybrid
stars only exist if a pure quark phase exists in the interior. We may expect
that the realistic situation lies between both descriptions, and, therefore,
we will analyze the implications of applying a Maxwell construction in the
next section.

In Fig.\ \ref{fig1} we show the  gravitational mass versus central pressure for various combinations of the 
three quantities $G_V,\, B^*,\, \sigma$. For a given EOS, these curves are obtained solving the well known Tolman-Oppenheimer-Volkov (TOV) \cite{shapiro83} 
equations describing the hydrostatic equilibrium general relativity. Hadronic star sequences are calculated using the NL3 parametrization 
considering pure nucleonic matter (black curve). The hybrid star (YS) sequence is represented by the dashed red curve.
%For the quark models considered in this paper, all QS sequences are made of hybrid stars (YS).  
%Results in the left (right) panel are relative to the NJL (CDM) model for the quark phase.  
The configuration marked with an asterisk represents, in all cases, the hadronic star for 
which the central pressure is equal to $P_0$ and thus the quark matter nucleation time 
is $\tau = \infty$.  The critical mass configuration is denoted by a full circle.  
The final conversion \cite{b0,b1,b2} of the critical mass configuration into a 
final quark star with the same stellar baryonic mass is denoted by a filled square.  
Notice that in most of the cases reported in the figures the quark matter nucleation process 
will lead to the formation of a black hole (BH). 

\begin{figure}%[b]
% \centering
\vspace{0cm}
% \begin{tabular}{cc}
\includegraphics[width=1.\linewidth,angle=0]{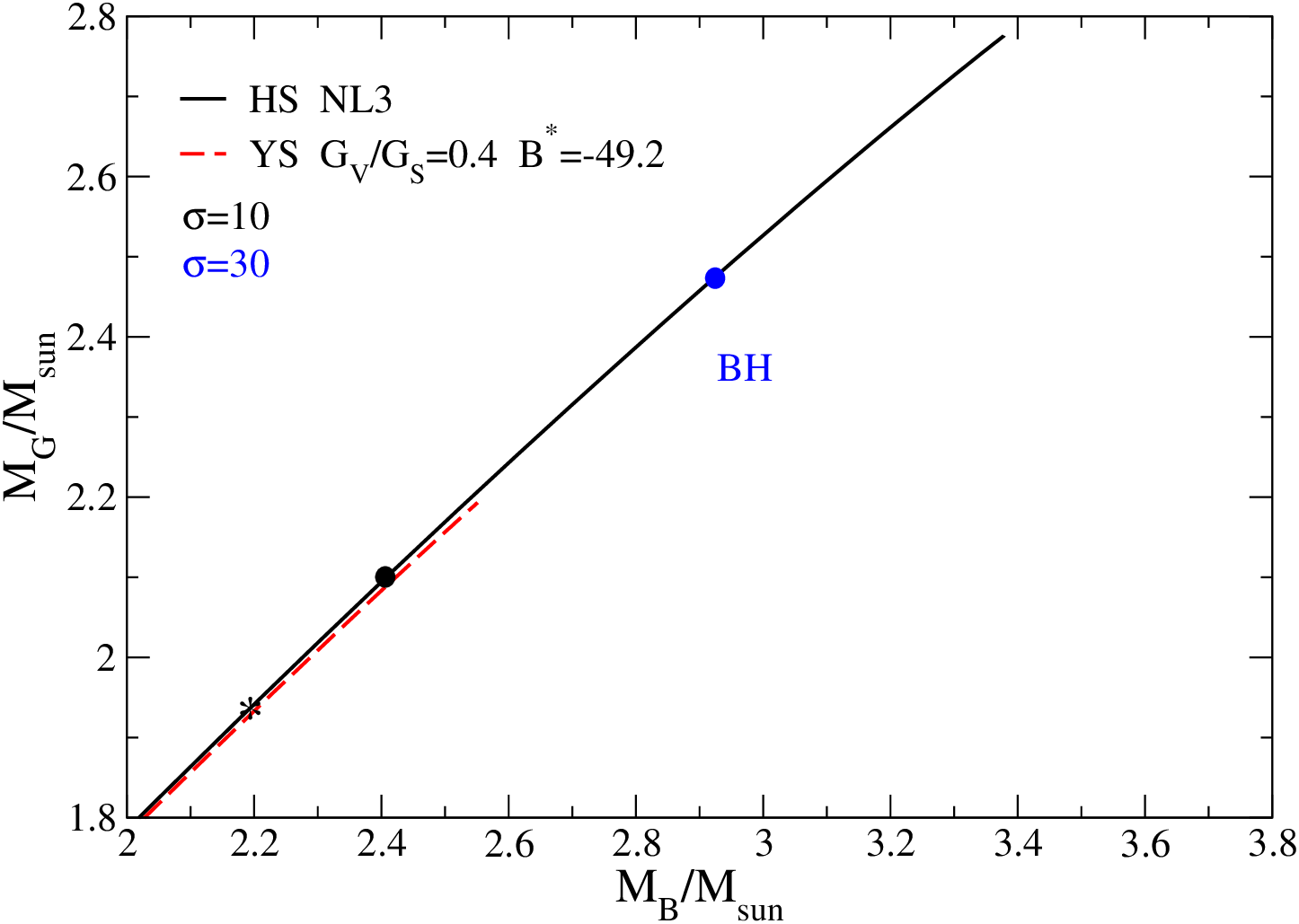}
% \end{tabular}
\caption{(Color online) Evolution of a hadronic star in the gravitational-baryonic mass plane using the NJL model with $G_V/G_S=0.4$ and $B^*=-49.2$ MeV/fm$^3$ 
to describe the quark phase, and the NL3 model for the hadronic phase.
The (black) line represents the cold  hadronic stars (HS) sequence. We consider two different values of the surface tension 
$\sigma=10, \ 30$ MeV/fm$^2$ at the interface between the hadronic and the quark phase. 
The asterisk and the full circle on these lines represent the stellar
configuration with nucleation time $\tau=\infty$ 
and the critical mass configuration $\tau=1$ yr, respectively. The lower red dashed line represents 
the cold YS sequence. 
Assuming $M_B$ constant, the evolution of a neutron star in this plane occurs along a vertical line.  
For $\sigma=30$ MeV/fm$^2$ the nucleation process leads to the formation of a black hole (BH). }
\label{fig2}
\end{figure}

In all panels of  Fig.\ \ref{fig1} the blue and black colors refer to the calculation in which the surface tension has been 
assumed equal to $\sigma=10$ MeV/fm$^2$ and $\sigma=30$ MeV/fm$^2$, respectively. In this calculation the strength 
of the vector interaction has been taken as $G_V/G_S=0.4$ while for the effective bag pressure $B^*$ we have set $B^*=-49.29$ MeV/fm$^3$ ( panels (b) and (d) ) and 
$B^*=-29.5$ MeV/fm$^3$ (panels (a) and (c)). For $\sigma=10$ MeV/fm$^2$ and $B^*=-49.29$ MeV/fm$^3$ a stable neutron star can be formed  after the nucleation process while, in all the other cases, the final configuration collapses into a black hole.  
The stable final star, obtained using the parameters discussed above,
is a neutron star with a pure quark content and not a simple hybrid
star with a mixed phase in its core.  
Similar results are shown in panels (c) and (d) for a calculation in which we put $G_V/G_S=0$ and we consider $B^*=0$ MeV/fm$^3$ (panel (c)) and 
$B^*=-29.5$ MeV/fm$^3$ (panel (d)). In this case all the final configurations are black holes. 
An equivalent way of presenting these results is shown in
Fig.\ \ref{fig2} where we plot the evolution of a hadronic star in the gravitational mass ($M_G$) versus baryonic mass($M_G$). In our calculation
we assume $M_B$ constant during the nucleation process, therefore the
evolution of neutron stars proceeds on a straight vertical line in
this plane. 

%%%%%%%%%%%%%%%%%%%%%%%%%%%%%%%%%%%%%%%%%%%%%%%%%%%%%%%%%%%%%%%%%%%%%%%%%%%%
% 
%
%%%%%%%%%%%%%%%%%%%%%%%%%%%%%%%%%%%%%%%%%%%%%%%%%%%%%%%%%%%%%%%%%%%%%%%%%%%%

 %%%%%%%%%%%%%%%%%%%%%%%%%%%%%%%%%%%%%%%%%%%%%%%%%%%%%%%%%%%%%%%%%%%%%%%%%%%%%%%%%%%%%%%%%%%%%%%%%%%%%%
\begin{table*} 
 \begin{center}                          
 \begin{ruledtabular}
 \begin{tabular}{llccccccccc}
 & $\sigma$ (MeV/fm$^2$) & $B^*$ (MeV/fm$^{3}$) & $P_0$ (MeV/fm$^{3}$)
 & $M (P_0)$ $(M_\odot)$ & $M_{cr}$ $(M_\odot)$ & $M^b_{cr}$ $(M_\odot)$ & $M_{fin}$ $(M_\odot)$ & $M_{max}^{YS}$ $(M_\odot)$ \\ 
%& $M_{max}^{QS}$ $(M_\odot)$ \\% &           & (MeV/fm$^{3}$) & $(M_\odot)$ &$(M_\odot)$ & $(M_\odot)$ &%     $(M_\odot)$  & $(M_\odot)$ \\
 \hline  \\
$G_V=0$       &  5 &     0       & 91.23  & 2.19   & 2.25 & 2.61 & BH    & 2.07 \\
              & 10 &     0       & 91.23  & 2.19   & 2.28 & 2.66 & BH    & 2.07 \\
              & 30 &     0       & 91.23  & 2.19   & 2.43 & 2.84 & BH    & 2.07 \\
              & 10 &   -29.59    & 38.98  & 1.59   & 1.79 & 2.00 & BH    & 1.75 \\
              & 30 &   -29.59    & 38.98  & 1.59   & 2.17 & 2.51 & BH    & 1.75 \\
              &     &        &          &          &        &        &        &   \\
$G_V/G_S=0.4$ 
              %& 5  &    29.59    &270.59  & 2.72   & 2.73 & 3.30 & BH    & 2.58 \\
              & 5  &     0       &212.75  & 2.65   & 2.66 & 3.21 & BH    & 2.50 \\ 
              & 30 &     0       &212.75  & 2.65   & 2.71 & 3.28 & BH    & 2.50 \\
              &  5 &   -29.59    &135.98  & 2.46   & 2.49 & 2.95 & BH    & 2.35 \\
              & 30 &   -29.59    &135.98  & 2.46   & 2.63 & 3.10 & BH    & 2.35 \\
              &  5 &   -39.46    & 97.87  & 2.27   & 2.33 & 2.73 & BH    & 2.27 \\
              & 30 &   -39.46    & 97.87  & 2.27   & 2.57 & 3.07 & BH    & 2.27 \\
%  NY          & 29 &-39.46    & 21.63  & 2.22   & 2.22 & 2.56 & 2.20  & 2.23\\
              &  5 &   -49.23    & 60.35  & 1.92   & 2.00 & 2.27 & 1.99  & 2.19 \\
              & 10 &   -49.23    & 60.35  & 1.92   & 2.09 & 2.40 & 2.08  & 2.19 \\
              & 15 &   -49.23    & 60.35  & 1.92   & 2.20 & 2.54 & 2.18  & 2.19 \\
              & 30 &   -49.23    & 60.35  & 1.92   & 2.45 & 2.89 & BH    & 2.19 \\ 
$G_V/G_S=0.4$ 
  NY          & 5  &   -39.46    & 189.93  & 2.30   & 2.31 & 2.70 & BH    & 2.23\\
$G_V/G_S=0.2$ 
  NY          & 9  &   -49.23    & 6.81    & 0.6    & 2.15 & 2.48 & 2.11  & 2.13\\
\end{tabular}
\end{ruledtabular}
\end{center} 
\caption{ The surface tension ($\sigma$), the parameter $B^*$, the transition
  pressure ($P_0$), the star mass with a central pressure equal to $P_0$
  ($M(P_0)$), the critical gravitational mass ($M_{cr}$) and baryonic mass
  ($M^b_{cr}$), the final mass ($M_{fin}$) and the maximum hybrid star mass
  ($M^{YS}$) obtained for the NL3 hadronic EOS including only nucleons in the
  hadronic phase except for the last two  lines identified with  `NY' which contain
  hyperons (see discussion in the text). The quark phase is described using the NJL model with and
  without vector interaction. The maximum quark star mass is the largest mass
  produced by integrating of the TOV equations and employing the EOS generated
  by the standard Gibbs construction. In this case the role of the surface
  tension between the hadronic and the quark phase is neglected.}
\label{tab_nl3}
\end{table*}
%%%%%%%%%%%%%%%%%%%%%%%%%%%%%%%%%%%%%%%%%%%%%%%%%%%%%%%%%%%%%%%%%%%%%%%%%%%%%%%%%%%%%%%%%%%%%%%%%%%%%%

In Table\ \ref{tab_nl3} we have reported the results of the
calculation of the nucleation process using the NL3 EOS, in particular
the following quantities are listed: surface tension (first column), $B^*$ (second column), $P_0$ (third column), mass of the neutron star with central pressure equal to $P_0$, $M(P_0)$ (fourth column), critical mass $M_{cr}$
(fifth column), critical baryonic mass $M^b_{cr}$ (sixth column), 
final mass $M_{fin}$ (seventh column) and maximum hybrid star mass $M_{max}^{YS}$. 

Combining the NJL model 
without vector interaction ($G_V=0$) with the NL3 EOS, we note that for $B^*=0$ and $B^*=-29.59$ MeV/fm$^3$ the 
nucleation process leads to the formation of black holes. When we include the vector interaction and we take the 
largest value of the effective bag constant considered $B^*=-49.23$ MeV/fm$^3$, we get stable final stars with mass
 compatible with the $2$ $M_\odot$ pulsar for value of the surface tension 
between $5$ and $15$ MeV/fm$^2$. As we have stated before, a negative
$B^*$ enlarges the quark content of the system while the vector
interaction goes in the opposite direction. A balance between these
two effects is needed in order to get stable final stars. The stable final 
stars obtained using the NL3 EOS are neutron stars with a pure quark
content. This calculation improves our previous results of Refs.\ \cite{b9} and \cite{me3} where we had found just low mass hybrid stars as final results of the nucleation process. 

However, it is worthwhile to note that although with this model the maximum hybrid star mass can be large ($2 M_\odot$ ) 
or very large ($2.50 M_\odot$) with $G_V=0$ or $G_V/G_S=0.4$ and $B^*\le
  -29.59$ MeV/fm$^3$,  the formation of such massive objects is not possible because the nucleation process leads always, in these cases, 
to a black hole. 
 The radius obtained for the $2.19 \ M_\odot$ neutron star is of
  $13.2$ Km. This value is just slightly out of the $M(R)$ constraints found
  in \cite{steiner13,steiner10}. Moreover, in  \cite{steiner13} constraints on the
  slope $L$ have also been imposed and for most of the models considered $L$
  should not exceed $65$ MeV. Both NL3 and TM1 have a quite high slope $L$
  (respectively, $118$ and $110$ MeV). Including a non-linear $\omega\rho$ term in
  the Lagrangian density it
  is possible to reduce $L$. A smaller $L$ will give rise to smaller stars
  still keeping almost unchanged the mass of the maximum mass configuration
  \cite{tm1-2}. All results shown in table \ref{tab_TM1}  were
  obtained with  TM1 and its modified TM1-2
including the $\omega\rho$ term for a symmetry energy slope
  $L=55$
  MeV. All radii are below $12.64$ km ($B^*=0$) and above $11.46$ km
  ($B^*=-39.46$)  in good agreement with the constraints of Steiner et al. \cite{steiner10,steiner13}.

Let us now discuss the results obtained with  TM1 and its modified TM1-2 which, as said before, satisfy the constraints obtained in
\cite{daniel02}, contrary to NL3. 
The parametrization TM1-2 has been chosen to be the hardest possible in
the range $2-3$ $\rho_0$ and still satisfy these constraints.
The results for these two models combined with the NJL model with $G_V=0$ and $G_V/G_S=0.2$ are summarized
in Table \ref{tab_TM1} . For $G_V=0$ there are several
combinations of parameters that allow the formation of stable hybrid stars,
but none of them is able to predict a star with a mass larger than $1.85$ $M_\odot$. The radius of this last configuration is 
$11.21$ km. 
  Including  the vector interaction in the
  NJL Lagrangian, taking $B^*=-39.46$ MeV/fm$^3$ and a value of the surface tension around 
$8$ MeV/fm$^2$, a hybrid star with mass of $2.03$ $M_\odot$ and a radius
of $12.41$ km is obtained. Just as for the NL3+NJL model, also with
the TM1-2+NJL model,  the quark vector interaction is essential to form a
stable high mass neutron star. 
For values of $G_V/G_S>0.2$ we cannot obtain any stable final star configuration using
the TM1 and the TM1-2 models. In this case the NJL model vector interaction is
so large that it pushes the quark onset to very large densities, inhibiting the nucleation process.    
  
This is indicative of how hard needs to be the hadronic EOS in order to
allow the formation of stars with a mass $\sim 2 M_\odot$ with a quark core. 
% Moving the upper limit of
%the constraints in \cite{daniel02} upwards by $\sim 10\%$ would be enough to re%produce $2$ $M_\odot$ stars. \\

In the last colum of table \ref{tab_TM1} we have reported both for TM1 and  TM1-2 models, the maximum neutron star mass obtained neglecting the finite surface effect at the interface between the hadronic and the quark matter. 
 For the TM1 and the TM1-2 models the largest masses
  resulted to be $M_G=2.00$ $M_\odot$ ($R=12.37$ km) and $2.07$
  $M_\odot$ ($R=12.65$ km), respectively.
These configurations can be populated if, after nucleation, the star goes through a process of mass accretion.

 The above masses have been produced setting $B^*=15.78$ MeV/fm$^3$ in both the cases. 
Larger neutron star masses can be obtained increasing the value of
$B^*$. However in all those cases 
the nucleation process leads to the formation of a black hole. 

\begin{figure}%[b]
 \vspace{0.5cm}
% \begin{tabular}{cc}
 \includegraphics[width=1.\linewidth,angle=0]{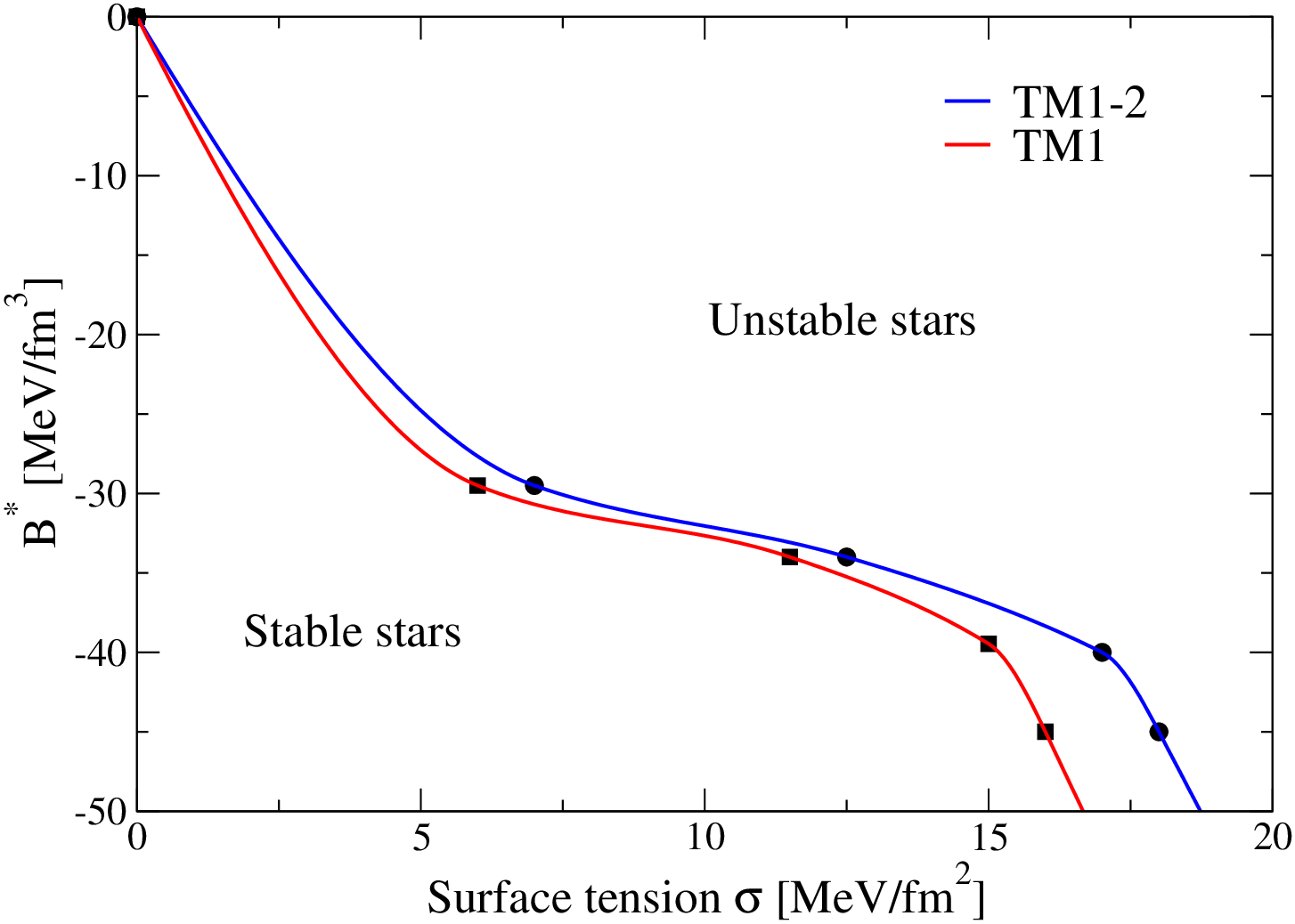}
% \end{tabular}
\caption{(Color online) 
The two curves represent the boundary of the region of parameters that allow for the formation of stable hybrid stars after the nucleation process. On the x-axis is reported the value of the surface tension $\sigma$ (in MeV/fm$^2$) while on the y-axis the value of $B^*$ 
(in MeV/fm$^3$).  Results are shown for the TM1 (red line) and the TM1-2 (blue
line) models. In the region below 
the  curves stable hybrid stars can be formed as a consequence of the nucleation process while, in the complementary region, nucleation  leads to the creation of black holes.}   
\label{fig3}
\end{figure}

Finally, in Fig.\ \ref{fig3} we have delineated the region of the parameters $B^*$ and $\sigma$ that allow for the formation of stable 
final stars after quark matter nucleation. Results are shown for the TM1 (red line) and the TM1-2 (blue line) models. Similar qualitatively results have been
obtained for the NL3 model, but we do not show them for simplicity. The circles and the squares in the figure have been obtained fixing 
for each value of $B^*$, the maximum $\sigma$ that allows for stable neutron stars after the nucleation process. 
This means that the combination of parameters that lie in the region under the curves leads to 
stable final stars while those in the complementary region to the formation of black holes. 
The effect of an effective 
bag pressure $B^*<0$ is to lower the onset of the quark phase. This
produces an enlargement of the window of metastable stars.  
In order to 
get stable final neutron stars, it is necessary to balance the effect of the
surface tension, that delays nucleation and allows for the creation of large massive
quark stars, and $B^*$, that tends to favor a nucleation at low pressures and densities 
reducing therefore the final maximum mass.

%%%%%%%%%%%%%%%%%%%%%%%%%%%%%%%%%%%%%%%%%%%%%%%%%%%%%%%%%%%%%%%%%%%%%%%%%%%%%%%%%%%%%%%%%%%%%%%%%%%%%%
\begin{table*} 
 \begin{center}
 \bigskip                           
 \begin{ruledtabular}
 \begin{tabular}{llccccccccc}
 Model & $\sigma$ (MeV/fm$^2$) & $B^*$ (MeV/fm$^{3}$) & $P_0$ (MeV/fm$^{3}$) & $M (P_0)$ $(M_\odot)$ & $M_{cr}$ $(M_\odot)$ & $M^b_{cr}$ $(M_\odot)$ & $M_{fin}$ $(M_\odot)$ & $M_{max}^{YS}$ $(M_\odot)$ \\ %& $M_{max}^{QS}$ $(M_\odot)$ \\
% &           & (MeV/fm$^{3}$) & $(M_\odot)$ &$(M_\odot)$ & $(M_\odot)$ &
%     $(M_\odot)$  & $(M_\odot)$ \\
 \hline  \\
TM1 ($G_V=0$)
& 5 &  15.78 & 257.16  & 2.11 & 2.12 & 2.48 & BH   & 2.00 \\
& 5 &   0    & 206.72  & 2.08 & 2.09 & 2.44 & BH   & 1.97 \\
& 5 & -15.78 & 147.46  & 1.99 & 2.02 & 2.34 & BH   & 1.92 \\  
& 6 & -29.59 &  82.89  & 1.75 & 1.83 & 2.10 & 1.83 & 1.84 \\ 
&10 & -29.59 &  82.89  & 1.75 & 1.88 & 2.16 & BH   & 1.84 \\
&15 & -39.46 &  26.48  & 1.10 & 1.79 & 2.03 & 1.78 & 1.80 \\
&20 & -39.46 &  26.48  & 1.10 & 1.93 & 2.23 & BH   & 1.80 \\
&16 & -45.00 &   6.90  & 0.48 & 1.86 & 2.13 & 1.83 & 1.88 \\
&20 & -45.00 &   6.90  & 0.48 & 2.02 & 2.35 & BH   & 1.88 \\
&     &       &  &  &  &  &   &   \\
TM1-2 ($G_V=0$)
& 5 &  15.78 & 206.24  & 2.18 & 2.19 & 2.58 & BH   &  2.07 \\ 
& 5 &   0    & 166.69  & 2.12 & 2.14 & 2.50 & BH   &  2.03 \\
& 5 & -15.78 & 120.17  & 2.00 & 2.03 & 2.36 & BH   &  1.96 \\ 
& 5 & -29.59 &  69.02  & 1.72 & 1.80 & 2.05 & 1.80 &  1.83 \\ 
& 7 & -29.59 &  69.02  & 1.72 & 1.83 & 2.08 & 1.82 &  1.83 \\
&10 & -29.59 &  69.02  & 1.72 & 1.87 & 2.14 & BH   &  1.83 \\
&10 & -39.46 &  24.99  & 1.09 & 1.56 & 1.73 & 1.55 &  1.80 \\
&17 & -39.46 &  24.99  & 1.09 & 1.81 & 2.06 & 1.80 &  1.80 \\
&20 & -39.46 &  24.99  & 1.09 & 1.90 & 2.18 & BH   &  1.80 \\
&18 & -45.00 &   6.89  & 0.48 & 1.88 & 2.15 & 1.85 &  1.88 \\
&20 & -45.00 &   6.89  & 0.48 & 1.97 & 2.27 & BH   &  1.88 \\
              &     &        &          &          &        &        &        &   \\
TM1-2 ($G_V/G_S=0.2$) 
& 5 & -39.46 & 155.88  & 2.10 & 2.14 & 2.51 & BH   &  2.06 \\
& 5 & -45.00 &  82.51  & 1.82 & 1.99 & 2.29 & 1.98 &  2.04 \\
& 8 & -45.00 &  82.51  & 1.82 & 2.04 & 2.37 & 2.03 &  2.04 \\
&10 & -45.00 &  82.51  & 1.82 & 2.07 & 2.42 & BH   &  2.04 \\ %(R_max=12.42 km)
 \end{tabular}
 \end{ruledtabular}
 \end{center} 
 \caption{ 
The surface tension ($\sigma$), the parameter $B^*$, the transition pressure
($P_0$), the star mass with a central pressure equal to $P_0$ ($M(P_0)$), the critical
gravitational mass ($M_{cr}$) and baryonic mass ($M^b_{cr}$), the final mass
($M_{fin}$) and the maximum quark star mass ($M^{YS}$) obtained for the  
TM1 and TM1-2 hadronic EOS including only nucleons and  including a $\omega\rho$ term so that the
slope of the symmetry energy is $L=55$ MeV.  The maximum hybrid star mass is the largest mass produced by integrating 
the TOV equations and employing the EOS generated by the standard Gibbs construction. In this case the role of the surface tension 
between the hadronic and the quark phase is neglected. }
 \label{tab_TM1}
 \end{table*}
%%%%%%%%%%%%%%%%%%%%%%%%%%%%%%%%%%%%%%%%%%%%%%%%%%%%%%%%%%%%%%%%%%%%%%%%%%%%%%%%%%%%%%%%%%%%%%%%%%%%%%
%
%We will now discuss the results obtained with the two microscopic hadronic EOS considered and, 
%in particular, we will comment whether the possible discussed scenarios are compatible with the
%recent measurement \cite{Demorest10} of the mass of the pulsar PSR J1614-2230  with a mass  
%M = (1.97 $\pm$ 0.04) $M_\odot$. 
%

%%%%%%%%%%%%%%%%%%%%%%%%%%%%%%%%%%%%%%%%%%%%%%%%%%%%%%%%%%%%%%%%%%%%%%%%%%%%%%
%%%%%%%%%%%%%%%%%%%%%%%%%%%%%%%%%%%%%%%%%%%%%%%%%%%%%%%%%%%%%%%%%%%%%%%%%%%%%%

%%%%%%%%%%%%%%%%%%%%%%%%%%%%%%%%%%%%%%%%%%%%%%%%%%%%%%%%%%%%%%%%%%%%%%%%%%%%%%

\section{Comments on the Maxwell construction and the inclusion of hyperons}

In this section we briefly discuss the dependence of the previous results on the approach used to construct the final neutron star
configurations which result from a nucleation process. All the calculations shown were performed
according to the Gibbs criterion. In the following,  we consider
another possible approach based on the Maxwell construction. In this case the
hadronic and the quark phases are connected by a region with constant pressure
leading, therefore, to a sharp phase transition. Within this phase transition construction possible existing hybrid
stars will always have a pure quark core, and, contrary to the Gibbs
construction,  central cores with  mixed
hadron-quark matter are excluded.

In table \ref{tab_max_gib} we compare the maximum neutron star masses and
radii obtained using the Maxwell and the Gibbs constructions. 
For the hadronic phase we have used the TM1-2 model while for  the quark phase we
have employed the NJL model without vector interaction ($G_V=0$). 
We note that the maximum mass produced considering both  possibilities are very similar although the 
corresponding radii can be quite different in some cases. However, we want to stress, that the results of the nucleation process 
discussed in the previous section for the TM1-2 and the TM1 models, are
affected only slightly by the choice of adopting the Gibbs instead of the Maxwell
construction. In fact, the nucleation process described above takes into account the surface
energy and, therefore, the lowest mass configurations of hybrid stars obtained
within the Gibbs construction will not be populated because their central
pressures lie below the pressure of the critical mass configuration.  

 The results 
obtained using the TM1 model are similar to those reported in table
\ref{tab_max_gib} and are not shown for brevity.  
   
Using the NL3 model for the hadronic phase, the hybrid stars produced within the Maxwell construction 
 get unstable for $B^*=0$. A similar result was obtained also in \cite{yasutake09} performing a Maxwell construction with the NJL model but using a different hadronic EOS. Putting $B^*=-29.59$ MeV/fm$^3$ and $G_V=0$ a stable hybrid star branch can be obtained. The maximum mass 
of this sequence is $1.75 \ M_\odot$ with a radius of $12$ km. In this case both the Maxwell  and the Gibbs construction give rise to the
same hybrid star maximum mass configuration.  

We have also studied the effect of including hyperons in the hadronic equation
of state. As referred above, the
hyperon interactions, in particular, the hyperon-hyperon one, is not well
constrained. In the following we consider a set of parameters which allows for
quite high star masses \cite{tm1-2}. We include the meson with hidden strangeness $\phi$ as in
\cite{schaffner2012,tm1-2}, we fix the $\omega$-vector meson couplings according the SU(6)
symmetry, the $\rho$-vector meson couplings according to the hyperon isospin
and we fit the couplings of the $\sigma$-scalar meson to the hypernuclear
potentials in nuclear matter, with $U_\Lambda=-28$ MeV, $U_\Sigma=30$ MeV,
$U_\Xi=18$ MeV. Taking {$B^*=-49.23$} MeV/fm$^3$, and {$\sigma= 9.0$} MeV/fm$^2$
we obtain the results shown in the last line of table \ref{tab_nl3}
with entry NY. In this particular case, it is possible to get a $2.11$
$M_\odot$ stable hybrid star 
after nucleation, which includes hyperons. However, as expected, the largest
hybrid star configuration is smaller when hyperons are included but not
necessary much smaller if enough repulsion between hyperons exists:
compare the maximum hybrid star  mass obtained with $B^*=-39.46$, 
MeV/fm$^3$, $G_V/G_S=0.4$, $\sigma=5$ MeV/fm$^2$  with and without hyperons, respectively, $2.23 M_\odot$
and $2.27$ $M_\odot$. 
 %%%%%%%%%%%%%%%%%%%%%%%%%%%%%%%%%%%%%%%%%%%%%%%%%%%%%%%%%%%%%%%%%%%%%%%%%%%%%%%%%%%%%%%%%%%%%%%%%%%%%%

%%%%%%%%%%%%%%%%%%%%%%%%%%%%%%%%%%%%%%%%%%%%%%%%%%%%%%%%%%%%%%%%%%%%%%%%%%%%%%%%%%%%%%%%%%%%%%%%%%%%%%
\begin{table*} 
 \begin{center}                          
 \begin{ruledtabular}
 \begin{tabular}{llcccccc}
 & $B^*$ (MeV/fm$^{3}$) & $M^{YS, Maxwell}_{max} \ (M_\odot)$ & $R^{YS, Maxwell}_{max}$ (km) & $M^{YS, Gibbs}_{max} \ (M_\odot)$ & $R^{YS, Gibbs}_{max}$ (km) \\
 \hline  \\
% &       &        &          &          &           \\
$\Lambda_\omega=0.03$ 
&15.78   &  2.15  &  12.60   &    2.07  &   12.65    \\
&  0.    &  2.10  &  12.72   &    2.03  &   12.64    \\
&-15.78  &  1.99  &  12.83   &    1.96  &   12.58    \\
&-29.56  &  1.83  &  11.98   &    1.83  &   12.00    \\
 &      &        &          &          &            \\
$\Lambda_\omega=0$ 
&15.78   &  2.15  &  13.27   &    2.01  &   13.19    \\         
&  0.    &  2.06  &  13.45   &    1.95  &   13.07    \\ 
&-15.78  &  1.88  &  13.53   &    1.86  &   12.68    \\
&-29.56  &  1.75  &  11.91   &    1.76  &   11.89    \\  
\end{tabular}
\end{ruledtabular}
\end{center} 
\caption{ The parameter $B^*$ (second column), the maximum hybrid star masses predicted by the Maxwell (third column) and the Gibbs (fifth column) 
construction. 
The corresponding radii are reported in columns fourth and sixth,
respectively. Results are shown for the TM1-2 hadronic EOS including only
nucleons in the hadronic phase with  ($\Lambda_\omega=0.03$) and without
($\Lambda_\omega=0$) the $\omega\rho$-term. The quark phase is described using the NJL model without vector interaction ($G_V=0$). The maximum quark star mass is the largest mass produced by integrating the TOV equations and employing the EOS. In this case the role of the surface tension between the hadronic and the quark phase is neglected.}
\label{tab_max_gib}
\end{table*}
%%%%%%%%%%%%%%%%%%%%%%%%%%%%%%%%%%%%%%%%%%%%%%%%%%%%%%%%%%%%%%%%%%%%%%%%%%%%%%%%%%%%%%%%%%%%%%%%%%%%%%  

%%%%%%%%%%%%%%%%%%%%%%%%%%%%%%%%%%%%%%%%%%%%%%%%%%%%%%%%%%%%%%%%%%%%%%%%%%%%%%%%%%

\section{Summary and Conclusions}
\label{sec:conclusions}

In this work we have analyzed the possibility of getting stable high mass
neutron stars, compatible with the recent observation of massive neutron stars,
as a consequence of a quark
matter nucleation process. We have considered three hadronic matter
EOS based on the RMF approach together with a three-flavor NJL model to
describe quark matter. The effect on the metastability of hadronic stars  of including a vector interaction
and  a phenomenological bag constant in the NJL model
 was discussed.

Using the TM1, TM1-2  and NL3 models to describe the hadronic phase,
we have shown that it is possible to obtain stable final stars after quark 
matter nucleation. In particular, in order to get
stable neutron stars, using the NL3 model, it is essential to include  
the vector term in the NJL Lagrangian density while, for the TM1 and
TM1-2 models, the stability of the final star configuration can only be
obtained with a weak or zero vector interaction.  For the TM1 model, we have obtained
slightly less massive stars than the ones predicted by TM1-2 one. 

We want to stress that the largest stable final mass obtained with the
NL3 model, namely $2.18$ $M_\odot$, is a neutron star containing a quark
core while the largest mass predicted by the TM1-2 model, that reads
$2.03 M_\odot$, is an hybrid star with a central core made of a mixed
phase. These values are both compatible with the mass of the pulsars
PSR J1614-2230 \cite{Demorest10}, $1.97\pm 0.04\, M_\odot$, and  PSR J0348+0432 \cite{j0348},
$2.01\pm 0.04\, M_\odot$. Note that if after nucleation the star suffers a long-term mass
accretion  from a companion star in
a binary system a star as massive as $2.04$ $M_\odot$ could be achieved within the TM1 parametrization.

According to the calculations performed in this work,
the location of the deconfinement phase transition in the phase 
diagram of
QCD, which  in our work, depends on the hadronic EOS used and the 
phenomenological bag
pressure $B^*$, plays  a very important role on the existence of quark matter in neutron 
stars.  
In addition the hadronic part of the system should be
sufficiently hard to preserve star stability. 
It was shown
   that not all massive quark star configurations are populated after
   nucleation. In particular, a too large surface tension may originate a
   black-hole after nucleation. In order to have conclusive results a
     study of the possibility that nucleation occurs at finite temperature
     should still be carried out. 

The vector interaction in the quark model allows
the formation of hybrid stars with a pure quark core, however, this is only
possible if the
hadronic EOS is very hard. In particular, using an EOS that at intermediate
densities  was designed to satisfy the upper limit of the constraints obtained in
\cite{daniel02}, it is possible  to obtain hybrid stars only if no  vector
interaction or just a weak one  is included in the NJL model. 

We conclude that more conclusive results depend on a better knowledge of 
a) the hadronic EOS at intermediate densities, b)  the surface energy of a quark
cluster in a hadronic matter background and c)  the hyperon interaction.  
 
%%%%%%%%%%%%%%%%%%%%%%%%%%%%%%%%%%%%%%%%%%%%%%%%%%%%%%%%%%%%
  
\section*{Acknowledgments}

This work has been partially supported by  by the initiative 
QREN financed by the UE/FEDER through the Programme COMPETE under 
the grant SFRH/BD/62353/2009 and the
projects PTDC/FIS/113292/2009
and CERN/FP/123608/2011; and by NEW COMPSTAR, a COST initiative.

%%%%%%%%%%%%%%%%%%%%%%%%%%%%%%%%%%%%%%%%%%%%%%%%%%%%%%%%%%%%

%%%%%%%%%%%%%%%%%%%%%%%%%%%%%%%%%%%%%%%%%%%%%%%%%%%%%%%%%%%%%%%%%%%%%
 
 \end{document}